\documentclass[aps,twocolumn,english,10pt,nofootinbib]{revtex4}

\usepackage{xcolor}
\usepackage{graphicx}
\usepackage{latexsym}
\usepackage{dcolumn}
\usepackage{bm}
\usepackage{slashed}

\usepackage{hyperref}
\usepackage{amsmath,amssymb,physics}
\usepackage{parskip}
\usepackage{adjustbox}
\usepackage{mathtools}
\usepackage{empheq}
\usepackage[normalem]{ulem}
\usepackage{verbatim}
\usepackage[most]{tcolorbox}
\hypersetup{colorlinks=true, linkcolor=magenta, filecolor=cyan, urlcolor=blue}
\usepackage{enumitem}

\newcommand{\mr}{\mathrm}

\newcommand{\bo}{\mathbf}
\newcommand{\gcoup}{g_{\phi \gamma}}
\newcommand{\tgcoup}{\Tilde{g}_{\phi \gamma}}

\newlength\dlf

\newtcolorbox{sidecalc}[1][]{
  breakable,
  colback=black!5,
  boxrule=0pt,
  sharp corners,
  title=Side calculation,
  coltitle=black,
  colbacktitle=black!5,
  fonttitle=\bfseries,
  titlerule=0pt,
  left=6pt, right=6pt, top=8pt, bottom=6pt,
  #1
}

\begin{document}

\title{Evolution of Cosmological Magnetic Fields from Ultralight Dark Matter}

\author{Abdias Aires}
\email{abdias.aires@unesp.br}
\affiliation{Department of Physics, McGill University, Montreal, QC H3A 2T8, Canada}
\affiliation{Institute for Theoretical Physics, Sao Paulo State University, R.~Dr.~Bento~T.~Ferraz~271,~Bl.~II, Sao Paulo 01140-070, SP, Brazil}

\begin{abstract}
    We investigate the possible evolution paths of cosmological magnetic fields generated from ultralight dark matter 
    after recombination. We consider a model in which the Chern-Simons coupling between the electromagnetic field tensor and an oscillating light pseudoscalar induces a tachyonic instability in the electromagnetic field, generating a magnetic field with a narrow peak at the critical resonant scale. By taking account of magnetohydrodynamic effects, we find that
    this peak is preserved through reionization until present cosmic time for a large region of parameter space, while turbulent decay beginning at generation
    constitutes a second large region. We also identify parameter values for which turbulent decay
    is delayed until reionization or the field remains in the overdamped regime from the time of its generation. From their imprints
    on large-scale structure and other cosmological observables, these evolution paths can be used to distinguish post-recombination from
    pre-recombination models of magnetogenesis. 
\end{abstract}

\date{\today}

\maketitle

\section{Introduction}\label{sec:intro}

The existence of magnetic fields on cosmological scales is suggested by
indirect evidence from the late universe. The nonobservation of $\gamma$-ray
cascade emission indicates the lower bound

\begin{equation}
\label{eq:B_bound}
    B(r) > 10^{-17} \, \mr{G}
\end{equation}

\noindent for length scales $r$ larger than $\sim 10^{-1}$ Mpc, rising as $r^{-1/2}$
on smaller scales \cite{Acciari_2023,Aharonian_2023}. That the bound holds in cosmic voids, distant from 
astrophysical sources, suggests a cosmological origin for the magnetic fields. 
In addition, the fields observed in galaxies and galaxy clusters may have been
amplified from primordial seeds~\cite{Kulsrud:1996km,Kulsrud:2007an,Brandenburg:2004jv,Widrow:2002ud}
(see~\cite{Grasso:2000wj,Kandus:2010nw,Durrer_2013,Subramanian:2015lua,Vachaspati:2020blt}
for reviews). These considerations have motivated various models of cosmological
magnetogenesis, usually active before Big Bang Nucleosynthesis, e.g., during inflation 
or phase transitions in the early universe \cite{Turner:1987bw,Sigl1997,Giovannini2008,Kahniashvili2010}.

Recently, a model of magnetogenesis based on ultralight dark matter (ULDM) has
been proposed to operate shortly after recombination \cite{brandenberger2025}. In this mechanism, a 
Chern-Simons coupling between the electromagnetic 
field tensor and a light pseudoscalar, assumed to compose most of dark matter, 
can induce a tachyonic instability in the electromagnetic field 
around the time of recombination. 
This process can potentially generate a strong magnetic field, with power initially
concentrated at the critical wavelength of the resonant process. For a large range
of coupling strengths, the initial coherence length is a sub-megaparsec
scale.

In this paper, we aim to distinguish the ULDM mechanism from pre-recombination magnetogenesis
by investigating the possible evolution paths of the magnetic field. We are especially interested
in the question of whether the narrow spectral peak at the critical resonant scale can be preserved until present cosmic time. 
The persistence of such a narrow peak, with potential imprints on structure formation and other cosmological 
observables, may be used to distinguish this model from pre-recombination magnetogenesis. Indeed, the latter typically
predicts an extended power-law (Kolmogorov) magnetic spectrum to be established by magnetohydrodynamic (MHD)
turbulence long before recombination. 

We approach this problem by determining the constraints in parameter space corresponding
to linear evolution and turbulent decay after recombination
and reionization. We identify the zones in parameter space corresponding to
four qualitative types of evolution: (1)~linear evolution ($\sim a^{-2}$) until present time; (2)~turbulent decay 
beginning when the magnetic fields are generated; (3)~turbulent decay beginning at reionization; (4)~Alfven 
overdamping until present time. The parameter space is dominated by cases (1) and (2). 
In cases (1) and (4), a narrow spectral peak at the critical wavenumber
is preserved until late times. In case (3), the peak is preserved until reionization, with
potential effects on structure formation permitting the model to be constrained. In case (2), the spectral
peak grows from sub-megaparsec to larger scales after recombination, a trajectory which may also 
distinguish this mechanism from pre-recombination magnetogenesis. These results can be extended to
other post-recombination models in which the generated magnetic field power is concentrated at a particular scale. 

This work is organized as follows. In section \ref{sec:magnetogenesis},
we review the ULDM model of magnetogenesis operating after recombination.
In section \ref{sec:MHD_decay}, we investigate the conditions
for linear evolution and MHD decay after recombination and reionization. In section \ref{sec:paths},
we determine the possible evolution paths of a magnetic field generated after recombination. Section
\ref{sec:conclusions} contains our conclusions.

Throughout this paper, we set $c = \hbar = 1$ and assume a $(-,+,+,+)$ metric signature.

\section{ULDM and magnetogenesis at recombination}\label{sec:magnetogenesis}

In this section, we review the ULDM model of magnetogenesis proposed 
in Ref. \cite{brandenberger2025}. It is based on the Lagrangian

\begin{equation}
    \mathcal{L} = -\frac{1}{2}\partial^\mu\phi \partial_\mu\phi
    - V(\phi) - \frac{1}{4}F_{\mu\nu}F^{\mu\nu} 
    + g_{\phi\gamma}\phi F_{\mu\nu} \Tilde{F}^{\mu\nu},
\end{equation}

\noindent where $\phi$ is a light pseudo-scalar field of potential
$V(\phi)$, coupled to the electromagnetic strength tensor $F_{\mu\nu}$
via a Chern-Simons term. The dual electromagnetic 
tensor is defined as $\Tilde{F}^{\mu\nu} \equiv \epsilon^{\mu\nu\rho\sigma}F_{\rho\sigma}/2$, 
where $\epsilon^{\mu\nu\rho\sigma}$ is the 4D Levi-Civita tensor. From this Lagrangian, we can
derive the following equation of motion of the gauge field:

\begin{equation}
\label{eq:hel_modes}
    (\partial^2 + k^2 \pm k g_{\phi\gamma}a(\eta)\dot\phi)\mathcal{A}_\pm = 0,
\end{equation}

\noindent where $\mathcal{A}_\pm$ are the two helical modes. This equation 
is valid after plasma effects are suppressed at recombination.\footnote{It is
an open question whether the residual conductivity in the post-recombination
universe can suppress the tachyonic instability described here. Whether this
effect is significant depends on the time-scale of the resonance as compared
to the mean time between collisions in the plasma. The reader is referred to Ref. 
\cite{Sharma2026} and the appendix of Ref. \cite{brahma2026} for a discussion.} The
field $\phi$ is 
assumed to have been
displaced from the minimum of its energy potential by some misalignment
mechanism in the early universe. It has begun to oscillate around that minimum
by the time of recombination. The equation above becomes tachyonic 
in the time-intervals for which $k \pm \gcoup a(\eta)\Dot{\phi} < 0$. 
Whenever this condition is satisfied, the corresponding helicity mode grows exponentially,
$\mathcal{A}_\pm \sim e^{2\mu_kt}$, where $\mu_k$ is the Floquet exponent

\begin{equation}
\label{eq:floquet}
    \mu_k = (g_{\phi\gamma} a m \phi(t)k)^{1/2}.
\end{equation}

\noindent As a result of the time-dependence of $a(\eta)$, the average growth of
the two helical modes is asymmetric, and the corresponding magnetic field will have
a small helicity.\footnote{This differs from helical generation during axion inflation, in which
only one helicity mode is unstable while the inflaton undergoes slow roll. The
resulting magnetic field is therefore strongly helical \cite{Durrer_2013}.} 
The maximal growth is obtained at the critical wavenumber

\begin{equation}
\label{eq:kc}
    k_c(\eta) \simeq g_{\phi\gamma} m \phi_0 a(\eta),
\end{equation}

\noindent which is the largest value for which \eqref{eq:hel_modes} can
be tachyonic. Here, we have used $\Dot{\phi} \sim m\phi_0$, where
$\phi_0$ is the amplitude of oscillations of the field $\phi$. One has
$m\phi_0 \sim T^2_\mr{rec}$ if $\phi$ composes
most of dark matter.\footnote{The energy density of dark
matter is $\sim T_\mr{rec}^4$ at recombination.} The critical wavenumber
can therefore be written as

\begin{equation}
\label{eq:lambda_c}
    k_c \simeq \Tilde{g}_{\phi\gamma} 10^{6} \, \mr{Mpc}^{-1},
\end{equation}

\noindent where $\Tilde{g}_{\phi\gamma}$ is defined as

\begin{equation}
    g_{\phi\gamma} \equiv \Tilde{g}_{\phi\gamma} 10^{-10} \, \mr{GeV}^{-1}.
\end{equation}
A conservative upper bound is 
\cite{Marsh_2016}

\begin{equation}
    \Tilde{g}_{\phi\gamma} \ll 1.    
\end{equation}

\noindent Because the phase space increases as $\sim k^3$ and the growth
of the magnetic field
is exponential in $\mu_k \sim k^{1/2}$, most of its power
will be concentrated at wavenumbers close to $k_c$.
The magnetic energy density will be written as $\rho_B = \mathcal{F}
\rho_\mr{cdm}$, where $\mathcal{F}$ is the fraction of energy density
drawn from dark matter (of energy density $\rho_\mr{cdm}$) at generation.

\section{Conditions for MHD decay after recombination}
\label{sec:MHD_decay}

In this section, we investigate the conditions for MHD decay of a magnetic
field generated after recombination, including the period after reionization. 
We derive parameter constraints on the ULDM model corresponding to
linear evolution, MHD turbulence, and MHD overdamping. In section \ref{sec:paths},
we will use these constraints to identify the possible evolution paths of a magnetic
field generated by the ULDM mechanism.

The ionization fraction of the post-recombination plasma decreases from $\sim 1$ at
$z \simeq 1100$ to 
$\sim 10^{-1}$ by $z \simeq 1000$ and to $\sim 10^{-4}$ by $z \simeq 100$
\cite{sethi2005}.
As the parametric resonance of the ULDM model requires
near-vacuum conditions, we will often specify $z_i \simeq 100$ as the time of 
magnetogenesis, while allowing $z_i$ (or $a_i = (1+z_i)^{-1}$)
to be a free parameter in what follows.
We assume a Gaussian, isotropic, and homogeneous 
magnetic field, defining its comoving power spectrum through

\begin{equation}
\label{eq:power_spec}
    \langle \Tilde{B}_i(\bo q) \Tilde{B}^*_j(\bo k) \rangle
        = \delta^3_D(\bo q - \bo k) (\delta_{ij} - k_i k_j/k^2) P_B(k),
\end{equation}

\noindent where $\Tilde{\bo B}_i(k)$ is the Fourier transform of the comoving
field. 
In the linear regime, the physical field redshifts as $\bo{B} (\bo x,t)
\simeq \Tilde{\bo B}(\bo x)/a^2$. We denote as $\Tilde B_{(s)}(k)$ the comoving field
smoothed on a scale $k$,

\begin{equation}
\label{eq:comov_B}
    \Tilde{B}_{(s)}^2(k) \equiv \int_0^k dk' \, 8\pi k'^2 P_B(k').
\end{equation}

\noindent From \eqref{eq:power_spec}, the squared rms value of the magnetic field 
in real space is $\langle \Tilde{\bo B}^2(\bo x) \rangle = \int dk 8\pi k^2
P_B(k)$. We
approximate the initial spectrum as highly peaked at the
critical wavenumber of parametric resonance,

\begin{equation}
\label{eq:P_Dirac}
    P_B(k) = \frac{B_0^2}{8\pi k^2}  \delta(k-k_c).
\end{equation}

\noindent With the above normalization, the rms field strength is simply
$B_0$. (This approximation is justified in section \ref{sec:magnetogenesis}.)
For this spectrum, the comoving field smoothed on the scale $k$ is, from
the definition \eqref{eq:comov_B}

\begin{equation}
    \Tilde{B}_{(s)}(k) = B_0 \Theta(k-k_c).
\end{equation}

\noindent If the magnetic field draws an energy fraction $\mathcal{F}$ from
dark matter at generation, the condition $\langle B^2(\bo x,t) \rangle
= 8\pi \rho_B (t) = 8\pi \mathcal{F} \rho_\mr{cdm} (t)$ gives

\begin{equation}
\label{eq:B0}
    B_0 = \left[ 8\pi a_i\mathcal{F} \rho_c \Omega_c^{(0)} 
        \right]^{1/2}
        = 2.3 \times 10^{-4} (\mathcal{F} a_i)^{1/2} \,\mr{G},
\end{equation}
where $a_i$ is the scale factor at generation.

\subsection{Nonlinear processing}

Nonlinear processing
occurs
at all scales for which $v(l)/l \geq H(t)$, where $v(l)$ is the 
fluid velocity induced by the field at scale $l$ \cite{sethi2005}. For larger
scales, the magnetic field evolves in the linear 
regime (without mode coupling), $B \sim a^{-2}$. If the fluid's
Reynolds number is large, the velocity $v(l)$ is of the order of the Alfven velocity, 
$v(l) \approx V_A(k = a/l, t)$, given in the post-recombination
epoch by \cite{sethi2005}

\begin{equation}
\begin{aligned}
\label{eq:VA}
    V_A(k,t) &= \frac{B_{(s)}(k,t)}{[4\pi \rho_b(t)]^{1/2}} \\
        &\approx 1.5 \times 10^{-5} \frac{B_{(s)}(k,t)a^2}{10^{-9} \mr{G}}
        a^{-1/2},
\end{aligned}
\end{equation}

\noindent where $\rho_b(t)$ is the density of baryons and $B_{(s)}(k,t)$
is the physical magnetic field smoothed on a scale $l = a/k$. 

We first derive a condition for the magnetic field generated by the ULDM model
to be in the nonlinear regime at scale $k_c$. Given
$v(l) \approx V_A(k = a/l, t)$, the condition 
$v(l)/l \geq H(t)$ for $l = a/k_c$ can be written as

\begin{equation}
\begin{aligned}
\label{eq:nonlin}
    \frac{k_c}{a} V_A(k_c,t) &\geq H_0 \sqrt{\Omega_m} a^{-3/2} \\
    \frac{k_c}{a} \left[ 1.5 \times 10^{-5} \frac{B_0}{10^{-9}\mr{G}} 
    \right] a^{-1/2} &\geq
        (1.26 \times 10^{-4}) a^{-3/2},
\end{aligned}
\end{equation}

\noindent where we have approximated $B_{(s)}(t) \simeq B_0 a^{-2}$ at the limit
between linear and nonlinear processing. We use $t$ and $a$ interchangeably
as the time coordinate. With the normalization
\eqref{eq:B0} and $k_c \simeq \tgcoup 10^6 \,\mr{Mpc}^{-1}$, the Alfven velocity at the
scale $k_c$ is

\begin{equation}
\label{eq:VA_F}
    V_A(k_c, a) \simeq 3.4 \left( \frac{\mathcal{F}a_i}{a} \right)^{1/2},
\end{equation} 
and \eqref{eq:nonlin} translates to

\begin{equation}
\label{eq:nonlin_cond}
    \tgcoup (\mathcal{F}a_i)^{1/2} \gtrsim 3.7 \times 10^{-11}
\end{equation}

\noindent as the condition for nonlinear processing of the scale
$k_c$. For $a_i = 10^{-2}$, this is

\begin{equation}
\label{eq:nonlin_cond_102}
    \tgcoup \mathcal{F}^{1/2} \gtrsim 3.7 \times 10^{-10}, 
    \quad\quad a_i = 10^{-2}.
\end{equation}

We have assumed that ions and neutrals are driven as a single fluid by the
magnetic field after recombination. Before reionization, this assumption may not hold for all field strengths
and coherence scales studied in this paper. An important next step would be the verification
of these parameter boundaries by MHD simulations that incorporate a two-fluid model (see section
\ref{sec:paths}).

In the region of parameter space $(\tgcoup, \mathcal{F}, a_i)$
for which \eqref{eq:nonlin_cond} is not satisfied (under the assumption of a single fluid), 
the magnetic field
evolves without mode coupling and scales with $\sim a^{-2}$.
(The field strength decreases
somewhat more quickly than $a^{-2}$ as a result of ambipolar diffusion \cite{sethi2005}.) 
Note that \eqref{eq:nonlin_cond} is time-independent, valid as well
after reionization. It follows that the original spectral shape, with
a narrow peak at $k_c \simeq \tgcoup10^6 \, \mr{Mpc}^{-1}$, persists 
until $z \simeq 0$ in the parameter region for which \eqref{eq:nonlin_cond}
is not satisfied.

\subsection{MHD decay between recombination and reionization}

Turbulence will be established on nonlinear scales for which the
kinetic Reynolds number, $R_K(k)$, is larger than order unity,

\begin{equation}
\label{eq:Rk}
    R_K(k,t) = \frac{v(l \simeq a/k)}{k \Tilde{\nu}} \approx 
        \frac{V_A(k,t)}{k \Tilde{\nu}}.
\end{equation}

\noindent We again take $v(l \simeq a/k)$ to be of the order of the Alfven velocity,
\eqref{eq:VA} and \eqref{eq:VA_F}. The quantity $\Tilde{\nu}$ is a rescaled shear viscosity,
$\Tilde{\nu} \equiv \nu/a$ \cite{Durrer_2013}. It is determined by the species with the longest mean free time
between collisions, weighted by its abundance. Between recombination and reionization, viscosity results from
neutral hydrogen collisions \cite{Banerjee_2004},

\begin{equation}
\label{eq:visc_rec}
    \Tilde{\nu} \simeq 3.8 \times 10^{-12} \left( \frac{a}{10^{-2}} \right) \, \mr{Mpc}
\end{equation}

\noindent for $z \lesssim 150$. The scaling with $a$ follows from the adiabatic cooling
of baryons, $T_b \sim a^{-2}$. Turbulent MHD processing of the
scale $k_c$ requires $R_K(k_c, t) > 1$. From \eqref{eq:VA_F},

\begin{equation}
\label{eq:Reynolds_HH}
    \mathcal F^{1/2} a_i^{1/2} a^{-3/2} \tgcoup^{-1} \gtrsim 1.1\times10^{-4},
\end{equation}
valid for $a \gtrsim (150)^{-1}$. The strongest requirement 
is obtained at the edge of reionization ($a \simeq 10^{-1}$), assuming the same expression for the 
viscosity,

\begin{equation}
\label{eq:Reynolds_HH_reio}
    \mathcal F^{1/2} a_i^{1/2} \tgcoup^{-1} \gtrsim 3.5\times10^{-6}.
\end{equation}

Dissipation of energy at small scales causes the overall coherence length of the 
magnetic field to increase. Its rate of change is of the order of 
the Alfven velocity \cite{Durrer_2013}. The largest scale processed by turbulence
can be estimated as

\begin{equation}
\begin{aligned}
\label{eq:largest_lambda}
    \lambda_B(a) &\simeq \int_{\eta_i}^{\eta} d\eta' \, V_A(k_c,\eta')
        \\&= \int_{a_i}^{a} \frac{da'}{a'^2 H(a')} \, V_A(k_c,a')
        \\&= \frac{3.4\,(\mathcal{F}a_i)^{1/2}}{H_0 \sqrt{\Omega_m}}
            \ln \frac{a}{a_i}
        \\&\simeq 2.7 \times 10^4 \, (\mathcal{F}a_i)^{1/2}
            \ln \frac{a}{a_i}\ \mathrm{Mpc},
\end{aligned}
\end{equation}

\noindent where we have used \eqref{eq:VA_F} and neglected $\Lambda$ domination at late times. 

To obtain an
analytical estimate for $\lambda_B(a)$, we have evaluated $V_A$ at the initial 
coherence scale $k_c$ and assumed negligible energy loss due to MHD processing and ambipolar
diffusion. MHD simulations suggest that dissipation causes a slow, 
logarithmic decay of the field strength beyond the scaling $\sim a^{-2}$, under matter 
domination \cite{trivedi2018}. 
This effect is therefore less significant here than for models that generate a magnetic field during
radiation domination.\footnote{We also neglect the small
helicity arising from the asymmetry of the resonant ULDM process in helical modes.} We discuss these assumptions in section
\ref{sec:conclusions}.

\begin{figure*}[t]
    \centering
    \includegraphics[width=\textwidth]{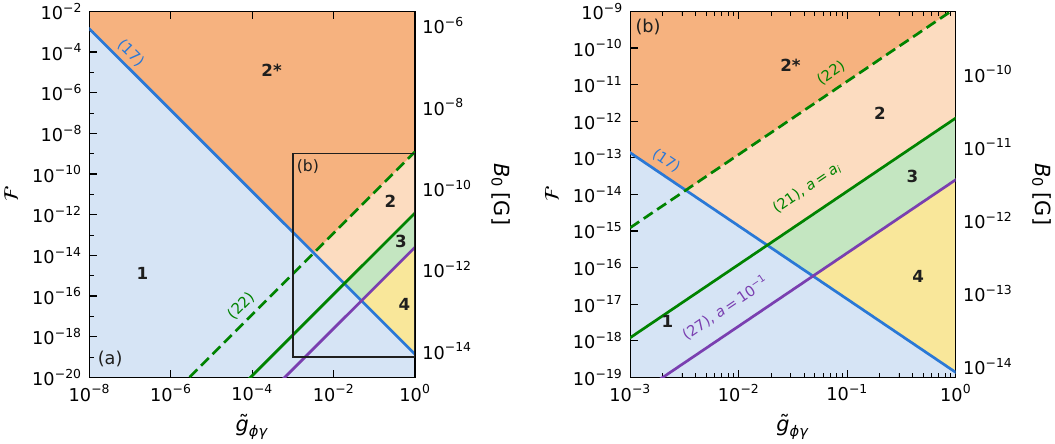}
    \caption{Evolution paths as a function of $\mathcal{F}$ and $\tgcoup$ for a magnetic field
    generated at $a_i = 10^{-2}$. Panel (b) is an enlargement of the region indicated in panel (a).
    The right-hand axis $B_0$ gives the comoving magnetic field strength corresponding to $\mathcal{F}$
    at generation. The numbered regions follow section \ref{sec:paths}:
    (\textit{1}) linear evolution; (\textit{2} and \textit{2}*) MHD turbulence beginning at
    generation (\textit{2}*: uninterrupted until today); (\textit{3}) MHD
    turbulence beginning at reionization; (\textit{4}) overdamped regime 
    since generation. The dividing lines are: the limit of linear--nonlinear evolution
    (solid blue); Reynolds number of order unity at $a \simeq 10^{-1}$, for viscosity governed by
    H--H collisions (dashed green); Reynolds number of order unity at $a=a_i$, for the same
    viscosity (solid green); Reynolds number of order unity at $a \simeq 10^{-1}$, 
    for viscosity governed by the Coulomb scattering of protons (purple).}
    \label{fig:paths}
\end{figure*}

\subsection{MHD decay after reionization}

At reionization, kinetic diffusion remains the main process of damping. Viscosity is now
dominated by proton-proton scattering in Coulomb collisions~\cite{Ghosh_2026},

\begin{equation}
    \nu = \frac{3 T_b^2}{4\sqrt{2\pi} e^4 \ln \Lambda_c n_b X_e(T)}
        \sqrt{\frac{T_b}{m_p}},
\end{equation}
where $n_b$ is the baryon number density, $X_e(T) \simeq 1$ is
the ionization fraction, $m_p$ is the proton mass, and $\ln \Lambda_c$ is the
Coulomb logarithm. The comoving viscosity $\Tilde{\nu} = \nu/a$ can therefore
be approximated as

\begin{equation}
\begin{aligned}
\label{eq:rescaled_nu_reio}
    \Tilde{\nu} &\simeq1.7\times10^{-11}\left(\frac{T_b}{10^4\,\mathrm K}\right)^{5/2}
    \left(\frac{\Omega_b}{0.049}\right)^{-1}
    \\&\quad\quad a^2\,X_e^{-1}
    \left(\frac{\ln\Lambda_c}{25}\right)^{-1}\ \mathrm{Mpc}.
\end{aligned}
\end{equation}

Two situations are especially relevant:
\begin{enumerate}[label=(\roman*)]
    \item MHD decay begins between recombination
    and reionization. To determine whether this process can continue after reionization,
    we must evaluate the kinetic Reynolds number at the scale $\lambda_B$ estimated
    in \eqref{eq:largest_lambda}, for $a \gtrsim 10^{-1}$. (For simplicity, we suppose the
    plasma to be fully reionized by $a \simeq 10^{-1}$.)
    \item The magnetic field is in the overdamped regime between recombination and reionization.
    Since the energy loss is approximately scale-independent, its spectrum 
    retains a peak at the original coherence scale $k_c$. The condition for MHD turbulence after
    reionization depends on the kinetic Reynolds number evaluated at $k_c$. 
\end{enumerate}

From \eqref{eq:VA_F}, \eqref{eq:Rk}, and \eqref{eq:rescaled_nu_reio}, and for $k_B \simeq 1/\lambda_B$ 
given by \eqref{eq:largest_lambda}, the requirement in case (\textit{i}) is

\begin{equation}
\label{eq:Reynolds_Coulomb_kB}
    R_K(k_B,a) \simeq5.4\times10^{15}\,\mathcal F a_i\,a^{-5/2}\ln\frac{a}{a_i} > 1.
\end{equation}

\noindent The condition in case (\textit{ii}) is

\begin{equation}
\label{eq:Reynolds_Coulomb_kc}
    R_K(k_c,a)\simeq2.0\times10^{5}\, (\mathcal F a_i)^{1/2} 
        \tilde g_{\phi\gamma}^{-1} \,a^{-5/2} > 1.
\end{equation}

\section{Evolution of the magnetic field}
\label{sec:paths}

From the results of the previous section, we can 
distinguish four qualitative cases in the evolution of the magnetic field 
generated after recombination. 
These are illustrated in Fig. \ref{fig:paths} for $a_i = 10^{-2}$.

\medskip

\noindent 1) \textit{Linear evolution}. For large enough scales
or small enough field strengths, the Hubble rate exceeds the rate
of energy transfer between modes by MHD processing.
This situation corresponds to parameter values that
violate \eqref{eq:nonlin_cond}. 
The magnetic field evolves without mode
coupling, redshifting as $\sim a^{-2}$. 
A sharp peak at $k_c \simeq \tgcoup 10^6 \mr{Mpc}^{-1}$
is therefore retained until today. The present rms field strength is 
(discounting the losses by ambipolar diffusion)
$2.3 \times 10^{-4} (\mathcal{F} a_i)^{1/2} \,\mr{G}$ (see
\eqref{eq:B0}).

\medskip

\noindent 2) \textit{MHD turbulence begins at the time of magnetogenesis.}
The field is in the nonlinear turbulent regime at the time of generation. 
This requires \eqref{eq:nonlin_cond} and \eqref{eq:Reynolds_HH} at $a = a_i$. (The
latter equation is only valid for $a_i \gtrsim 150^{-1} $.) As
the Reynolds number only decreases with time before reionization, either MHD turbulence begins at generation
or it only begins at reionization or later. A special subcase is that of \textit{uninterrupted
MHD decay from generation}: \eqref{eq:Reynolds_Coulomb_kB} is now necessary for the
decay to continue through reionization. Note that \eqref{eq:Reynolds_Coulomb_kB} at
$a=1$ is implied by \eqref{eq:nonlin_cond} and \eqref{eq:Reynolds_HH_reio}. Hence, if the field is 
still turbulent at reionization, it remains turbulent 
until present cosmic time. (This subcase corresponds to region 2* in Fig.
\ref{fig:paths}.) The
present coherence length in this case cannot be larger than

\begin{equation}
\begin{aligned}
\lambda_B &\simeq 2.7\times10^{4}\,(\mathcal{F}a_i)^{1/2}\ln a_i^{-1}\ \mathrm{Mpc}
  \\&\simeq 1.2\times10^{4}\,\mathcal{F}^{1/2}\ \mathrm{Mpc}, \qquad a_i = 10^{-2},
\end{aligned}
\end{equation}
obtained from \eqref{eq:largest_lambda} at $a = 1$.

\medskip

\noindent 3) \textit{MHD turbulence begins at reionization.} The magnetic
field is in the nonlinear but overdamped regime until reionization,
after which it begins to evolve by turbulent decay. The conditions
are \eqref{eq:nonlin_cond}, a violation of \eqref{eq:Reynolds_HH} at $a = a_i$, and
\eqref{eq:Reynolds_Coulomb_kc} at $a = 10^{-1}$.

\medskip

\noindent 4) \textit{Overdamped regime from the time of magnetogenesis.} The 
field is in the nonlinear and overdamped regime until present cosmic time. The conditions are \eqref{eq:nonlin_cond}
and a violation of \eqref{eq:Reynolds_Coulomb_kc} at $a = 10^{-1}$, which implies a 
violation of \eqref{eq:Reynolds_HH} at
$a = a_i$ for $a_i \gtrsim 3\times10^{-3}$.

\medskip

The parameter space is dominated by linear evolution (region 1 of Fig. 
\ref{fig:paths}) and 
uninterrupted MHD turbulence from the time of magnetogenesis (region 2*).
We note that the boundaries between these regions will change with the
inclusion of dissipation effects and with a two-fluid treatment
of the post-recombination plasma (ions and neutrals coupled by collisions),
which can be done with MHD simulation codes
such as \textsc{Pencil Code}~\cite{PencilCode2021, Brandenburg2019} and
\textsc{MPI-AMRVAC}~\cite{Keppens2023,PopescuBraileanu2022}.

\section{Conclusion}
\label{sec:conclusions}

We have investigated the possible evolution paths of magnetic fields
produced from ultralight dark matter after recombination. The parameter
space is dominated by (\textit{1}) linear evolution and (\textit{2*}) uninterrupted MHD turbulence from the time
of magnetogenesis. For larger Chern-Simons couplings and smaller magnetic field 
energy densities, (\textit{3}) MHD turbulence can be delayed until reionization
or (\textit{4}) the field can remain in the overdamped regime from the time of its generation. 
In cases 1 and 4, the narrow spectral peak generated by
the resonant process of Ref. \cite{brandenberger2025} persists through
reionization until present cosmic time. These trajectories differ from 
those predicted from pre-recombination models of magnetogenesis (e.g., from 
inflation or primordial phase 
transitions~\cite{Turner:1987bw,Sigl1997,Giovannini2008,Kahniashvili2010}), in which MHD decay 
typically establishes a Kolmogorov spectrum on comoving megaparsec scales long before
recombination. 

As relevant further work, we note
the inclusion of dissipation effects
and a two-fluid treatment in MHD
simulations of the post-recombination universe. The same simulations can include
the ULDM mechanism of magnetogenesis itself, taking account of the back-reaction effects
which eventually shut off the resonance. Beyond the
accurate determination of parameter boundaries (and the confirmation of
the qualitative cases identified in the present work), 
an advantage of such a treatment would be the
reduction of the model's parameter space to only two parameters (the mass of the dark matter particle
and the Chern-Simons coupling strength), while the epoch of magnetogenesis $a_i$ and the fraction $\mathcal{F}$
of magnetic field energy would be determined dynamically from the model.

\acknowledgments
I thank Robert Brandenberger for discussions and comments on the draft, and I thank
the Department of Physics of McGill University for its hospitality. I have used 
Claude Opus 5.5 (Anthropic) for numerical computations, which I have revised, and to generate the figure.
My work is supported by funds from CAPES - 001 and CNPq, grant number 165240/2023-7.
\bibliography{ref}

\end{document}